\documentclass[prl,aps,twocolumn,showpacs]{revtex4}

\usepackage{pxfonts}
\usepackage{graphicx}
\begin{document}

\title{Electrostatic Conveyer for Excitons}

\author{A.\,G. Winbow, J.\,R. Leonard, M. Remeika, Y.\,Y. Kuznetsova, A.\,A. High, A.\,T. Hammack, L.\,V. Butov}
\affiliation{Department of Physics, University of California at San
Diego, La Jolla, CA 92093-0319, USA}
\author{J. Wilkes, A.\,A. Guenther, A.\,L. Ivanov}
\affiliation{Department of Physics and Astronomy, Cardiff University, Cardiff CF24 3AA, United Kingdom}
\author{M. Hanson, A.\,C. Gossard}
\affiliation{Materials Department, University of California at Santa Barbara, Santa Barbara, California 93106-5050}

\date{\today}

\begin{abstract}
\noindent
We report on the study of indirect excitons in moving lattices -- conveyers created by a set of AC voltages applied to the electrodes on the sample surface. The wavelength of this moving lattice is set by the electrode periodicity, the amplitude is controlled by the applied voltage, and the velocity is controlled by the AC frequency. We observed the dynamical localization-delocalization transition for excitons in the conveyers and measured its dependence on the exciton density and conveyer amplitude and velocity. We considered a model for exciton transport via conveyers. The theoretical simulations are in agreement with the experimental data.

\end{abstract}


\date{\today}

\maketitle

An indirect exciton is a bound pair of an electron and a hole confined in spatially separated layers. Due to their long lifetimes, indirect excitons can travel over large distances before recombination \cite{Hagn1995, Butov1998, Larionov2000, Butov2002, Voros2005, Ivanov2006, Gartner2006, Gartner2007, Vogele2009, Hammack2009}. Furthermore, indirect excitons have a built-in dipole moment $ed$, where $d$ is the separation between the electron and hole layers, so their energy can be controlled by voltage: an electric field $F_{\rm z}$ normal to the layers results in the exciton energy shift $U = e d F_{\rm z}$ ~\cite{Miller1985}. This gives an opportunity to create in-plane potential landscapes for excitons $U(x,y) = e d F_{\rm z}(x,y)$ by laterally modulated voltage $V(x,y)$. Excitons were studied in a variety of electrostatically formed potential landscapes: ramps \cite{Hagn1995, Gartner2006}, lattices \cite{Zimmermann1997, Zimmermann1998, Krauss2004, Hammack2006, Remeika2009}, traps \cite{Hammack2006, Huber1998, Chen2006, High2009nl, High2009prl}, and circuit devices \cite{High2007, High2008, Grosso2009, Kuznetsova2010}. In this paper, we present an excitonic conveyer -- moving lattice created by a set of AC voltages. The excitonic conveyer realizes controlled transport of excitons as charged coupled devices (CCD) realize controlled transport of electrons \cite{Smith2010}.

\begin{figure}
\begin{center}
\includegraphics[width=5.5cm]{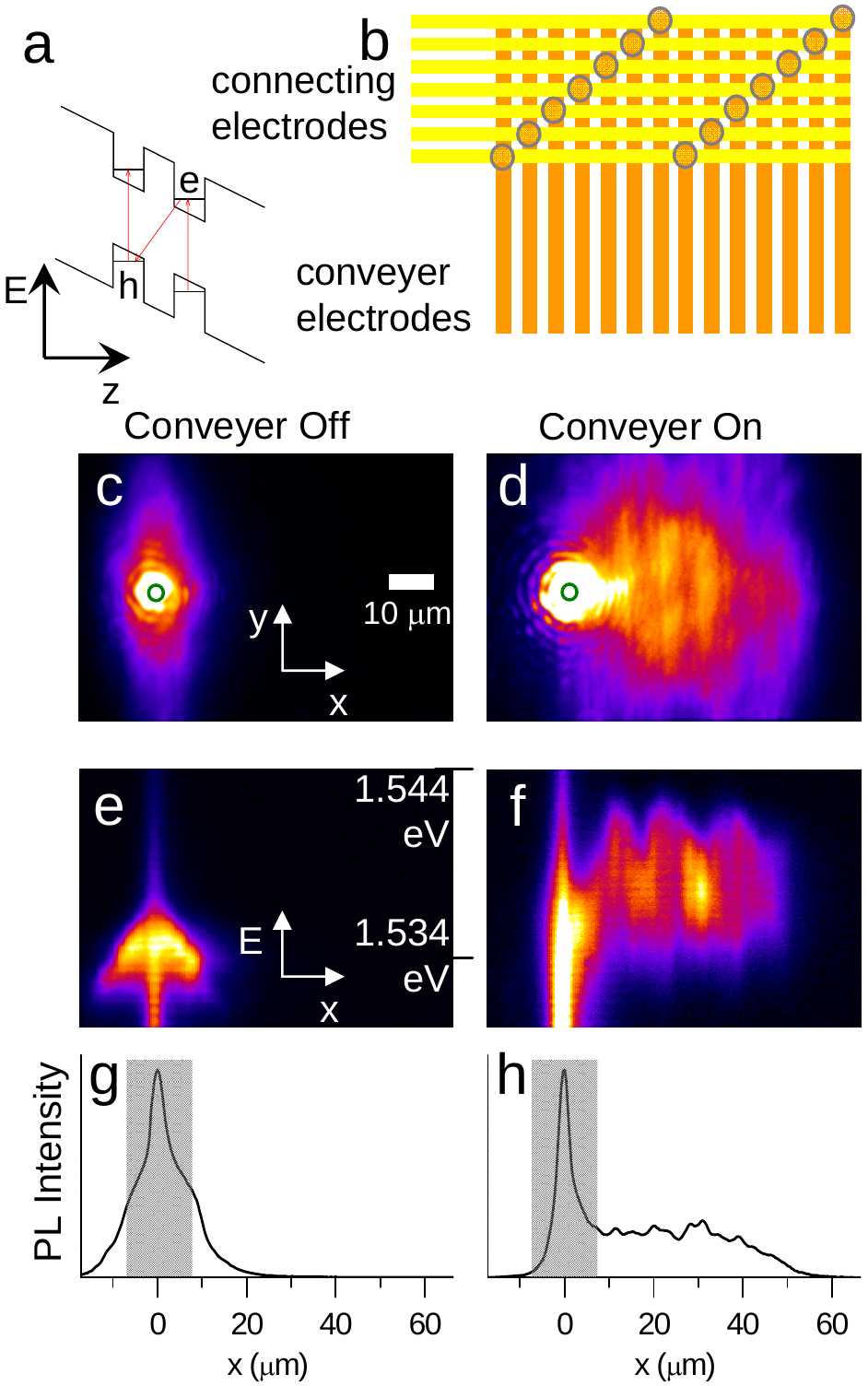}
\caption{(a) CQW band diagram; e, electron; h, hole. (b) Electrode pattern, contacts are shown by circles. (c-f) $x-y$ and $x-energy$ PL images for conveyer off and on. (g,h) PL intensity profiles $I(x)$. $P_{\rm ex} = 20\mu$W, $A_{\rm conv} = 7.5$ meV, $v_{\rm conv} = 0.7 \mu$m/ns.}
\end{center}
\end{figure}

Moving potential lattices can be created by surface acoustic waves (SAW). Transport of excitons, exciton-polaritons, and laterally separated electrons and holes via SAW is intensively studied \cite{Rocke1997, Rudolph2007, Lazic2010, Cerda2010}. The transport velocity in this case is defined by the sound velocity of SAW propagation $\sim 3 \mu$m/ns. In contrast, the velocity of the electrostatic excitonic conveyer can be controlled by the AC frequency and can be from well below to well above the sound velocity.

The indirect excitons are created in a GaAs coupled quantum well structure (CQW) grown by molecular beam epitaxy (Fig. 1a). An $n^+$-GaAs layer with $n_{\rm Si}=10^{18}$ cm$^{-3}$ serves as a homogeneous bottom electrode. Two 8 nm GaAs QWs separated by a 4 nm Al$_{0.33}$Ga$_{0.67}$As barrier are positioned 0.1 $\mu$m above the $n^+$-GaAs layer within an undoped 1 $\mu$m thick Al$_{0.33}$Ga$_{0.67}$As layer. Positioning the CQW closer to the homogeneous electrode suppresses the in-plane electric field \cite{Hammack2006}, which otherwise can lead to exciton dissociation \cite{Zimmermann1997}.

The conveyer potential is created by a set of semitransparent 1 $\mu$m wide 120 nm thick indium tin oxide (ITO) electrodes on the sample surface. The distance between the electrode centers is 2 $\mu$m, the conveyer periodicity is 7 electrodes, and the wavelength of the conveyer potential is $\lambda_{\rm conv} = 14 \mu$m. The conveyer electrodes are covered by a layer of transparent insulation ($300 \mu$m thick SiO$_2$). A set of connecting electrodes ($10 \mu$m wide 300 nm thick ITO) provides the contacts to the conveyer electrodes through $1 \times 10 \mu$m etched openings in the insulating layer (Fig. 1b). The conveyer length is $380 \mu$m, width is $80 \mu$m.

\begin{figure}
\begin{center}
\includegraphics[width=5cm]{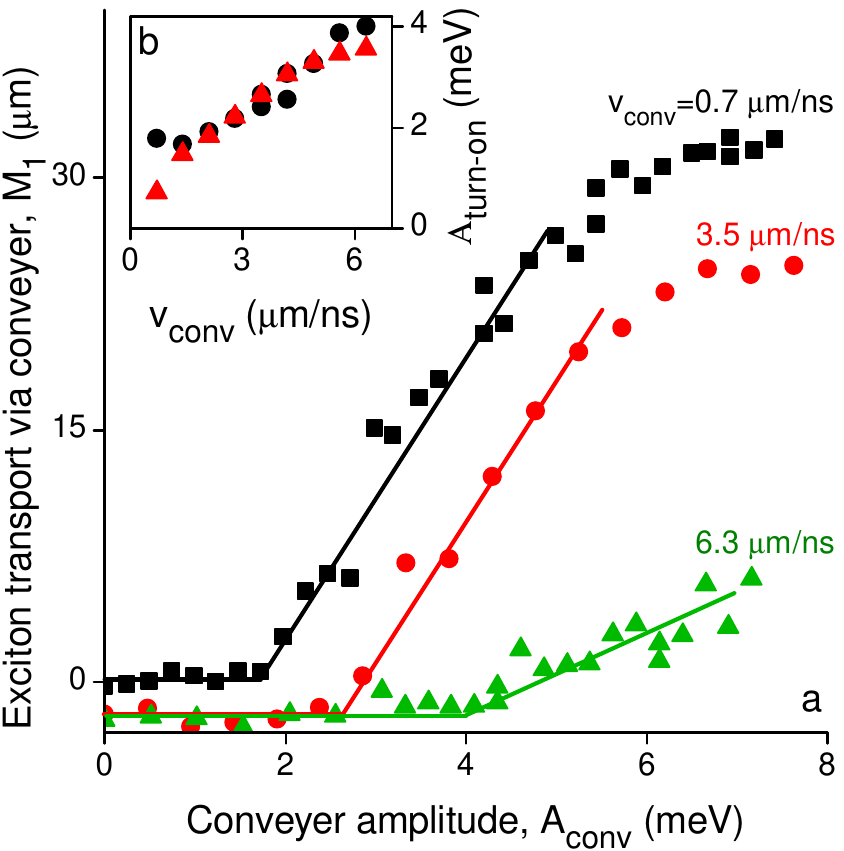}
\caption{(a) The average transport distance of indirect excitons via conveyer $M_1$ as a function of the conveyer amplitude, $A_{\rm conv}$. Lines are a guide to the eye, $A_{\rm turn-on}$ is $A_{\rm conv}$ at the line intersection. (b) The measured (black points) and calculated (red triangles) $A_{\rm turn-on}$ vs the conveyer velocity. $P_{\rm ex} = 20 \mu$W.}
\end{center}
\end{figure}

The sample mounts in a He cryostat at 1.7 K. AC voltages to the conveyer electrodes are applied by coaxial cables with impedance-matching termination at the sample. The regime, where the indirect excitons have lower energy than spatially direct excitons in the CQW, is realized by DC bias $V_{\rm bias}=4$ V supplied separately \cite{SOM}. A set of differentially phase-delayed AC sinewaves at frequency $f_{\rm conv}$ creates a traveling potential lattice for indirect excitons -- the excitonic conveyer. The amplitude of the conveyer potential for indirect excitons $A_{\rm conv}$ is controlled by the applied voltage \cite{SOM}. The conveyer velocity is controlled by the AC frequency $v_{\rm conv} = \lambda_{\rm conv} f_{\rm conv}$.

The excitons are photoexcited by 788 nm Ti:Sp laser focused to a spot $\sim 5 \mu$m in diameter. The exciton density is controlled by the laser excitation power $P_{\rm ex}$. Photoluminescence (PL) images of the exciton cloud are taken by a CCD with a bandpass filter $810 \pm 5$ nm covering the spectral range of the indirect excitons. The diffraction-limited spatial resolution is $1.4 \mu$m. The spectra are measured using a spectrometer with resolution 0.18 meV.

Figures 1c-f show $x-y$ and $x-energy$ PL images for conveyer off and on. The PL intensity profiles $I(x)$, obtained by the integration of the $x-energy$ images over the emission wavelength, are shown in Fig. 1g,h. Exciton transport via conveyer is presented by the extension of the exciton cloud along the direction of the moving potential. We quantify it by the first moment of the PL intensity $M_1 = \int x I(x) dx / \int I(x) dx$, which characterizes the average transport distance of indirect excitons via conveyer. The spectrally broad emission at $x=0$ (Fig. 1e,f) and sharp peak in $I(x)$ (Fig. 1g,h) originate from the bulk GaAs in the structure. To remove the contribution from the bulk, the shaded area is not included to the calculation of $M_1$ in the analysis of exciton transport.

Figure 2 presents exciton transport via conveyer as a function of the conveyer amplitude $A_{\rm conv}$. For a shallow conveyer, the exciton cloud extension $M_1$ is not affected by the conveyer motion indicating that the excitons do not follow the moving lattice, i.e. are dynamically delocalized in the lattice (Fig. 2a). In contrast, at higher conveyer amplitudes, excitons are moved by the moving lattice, i.e. are dynamically localized in the lattice. At the dynamical localization-delocalization transition (dLDT), the exciton cloud starts to follow the conveyer and $M_1$ changes from constant to increasing with $A_{\rm conv}$. We define the conveyer amplitude at the dLDT, $A_{\rm turn-on}$, as the point where the extrapolation of the growth of $M_1$ to small $A_{\rm conv}$ becomes equal to the low-$A_{\rm conv}$ constant. The dLDT is a dynamical counterpart of the LDT for excitons in static lattices \cite{Remeika2009}.

\begin{figure}
\begin{center}
\includegraphics[width=5cm]{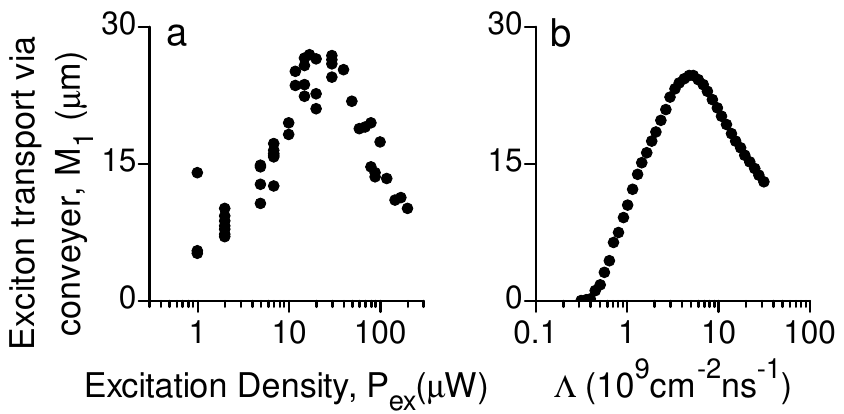}
\caption{(a) The measured and (b) calculated average transport distance of indirect excitons via conveyer $M_1$ as a function of density. $A_{\rm conv} = 4.9$ meV, $v_{\rm conv} = 0.7 \mu$m/ns.}
\end{center}
\end{figure}

The control of $f_{\rm conv}$ gives an opportunity to study exciton transport via conveyers and in particular the dLDT as a function of the conveyer velocity. Figure 2 shows that exciton transport via conveyer is less efficient for higher $v_{\rm conv}$. In particular, $A_{\rm turn-on}$ increases with $v_{\rm conv}$.

The control of $P_{\rm ex}$ gives an opportunity to study exciton transport via conveyers as a function of the exciton density. Figure 3a shows that excitons are hardly moved by the conveyer at low densities, efficient exciton transport via conveyer is achieved at intermediate densities, and exciton transport via conveyer becomes less efficient at high densities. The observed dependences of exciton transport via conveyer on the exciton density and conveyer velocity and amplitude are compared to the theoretical model and discussed below.

The following nonlinear partial differential equation was used to model in-plane transport of indirect excitons subject to the applied conveyer potential $U_{\rm conv}(x)$:
\begin {equation}
\frac{\partial n_{\rm x}}{\partial t} = \nabla \cdot \left[D_{\rm x} \nabla n_{\rm x} + \mu_{\rm x} n_{\rm x} \nabla(u_{\rm 0}n_{\rm x} + U_{\rm conv}) \right] + \Lambda - \frac{n_{\rm x}}{\tau_{\rm opt}}.
\label{transport}
\end {equation}
the first term in square brackets in Eq.\,(\ref{transport}) accounts for exciton diffusion, $D_{\rm x}$ is the diffusion coefficient. The second term accounts for exciton drift due to the dipole-dipole exciton interaction, which is approximated by $u_{\rm 0}n_{\rm x}$ \cite{Ivanov2002, Ivanov2010}, and the conveyer potential $U_{\rm conv} = e d F_{\rm z}(x) = e d \frac{\partial}{\partial z}V({\bf r})$, where voltage $V({\bf r})$ originates from the voltage applied to the conveyer electrodes $V_{\rm z=0}(x)$. The mobility $\mu_{\rm x}$ is given by the generalized Einstein relationship $\mu_{\rm x} = D_{\rm x}(e^{T_0/T} - 1)/(k_{\rm B} T_0)$, where $T_0=(2\pi\hbar^2 n_{\rm x})/(M_{\rm x} g k_{\rm B})$, $M_{\rm x} \simeq 0.22\,m_0$ is the exciton mass, $g=4$ is the spin degeneracy \cite{Ivanov2002}. Due to the geometry of the system, we use the approximation $\nabla = \partial/\partial x$ and solve for the density of indirect excitons $n_{\rm x}(x,t)$.

The effect of disorder intrinsic to QWs is included using a thermionic model for the diffusion coefficient, $D_{\rm x} = D_{\rm x}^{(0)}{\rm exp}\left(-U^{(0)}/(k_{\rm B}T + u_{\rm 0}n_{\rm x})\right)$ \cite{Ivanov2002}. $D_{\rm x}^{(0)}$ is the diffusion coefficient in the absence of QW disorder and $U^{(0)}/2$ is the amplitude of the disorder potential. The temperature of indirect excitons $T$ is approximated as $T = T_{\rm bath}$. The non-resonant photoexcitation causes heating of the exciton gas by a few Kelvin. However, the hot excitons cool to the lattice temperature within a few microns of the excitation spot \cite{Hammack2009} justifying the approximation.

The last two terms in Eq.\,(\ref{transport}) take account of the creation and decay of excitons. $\Lambda(x)$ is the generation rate and $\tau_{\rm opt}$ is the optical lifetime. The increased exciton velocity due to transport via the conveyer can, in principle, shift the energy of excitons outside the photon cone and increase their optical lifetime. However, we evaluated $\tau_{\rm opt}$ using the expressions given by Eqs. 1-6 in \cite{Hammack2009} and found that for the studied range of parameters, the corrections to the lifetime are small and can be neglected. Therefore we used a constant $\tau_{\rm opt}$ independent of $v_{\rm conv}$.

\begin{figure}
\begin{center}
\includegraphics[width=5cm]{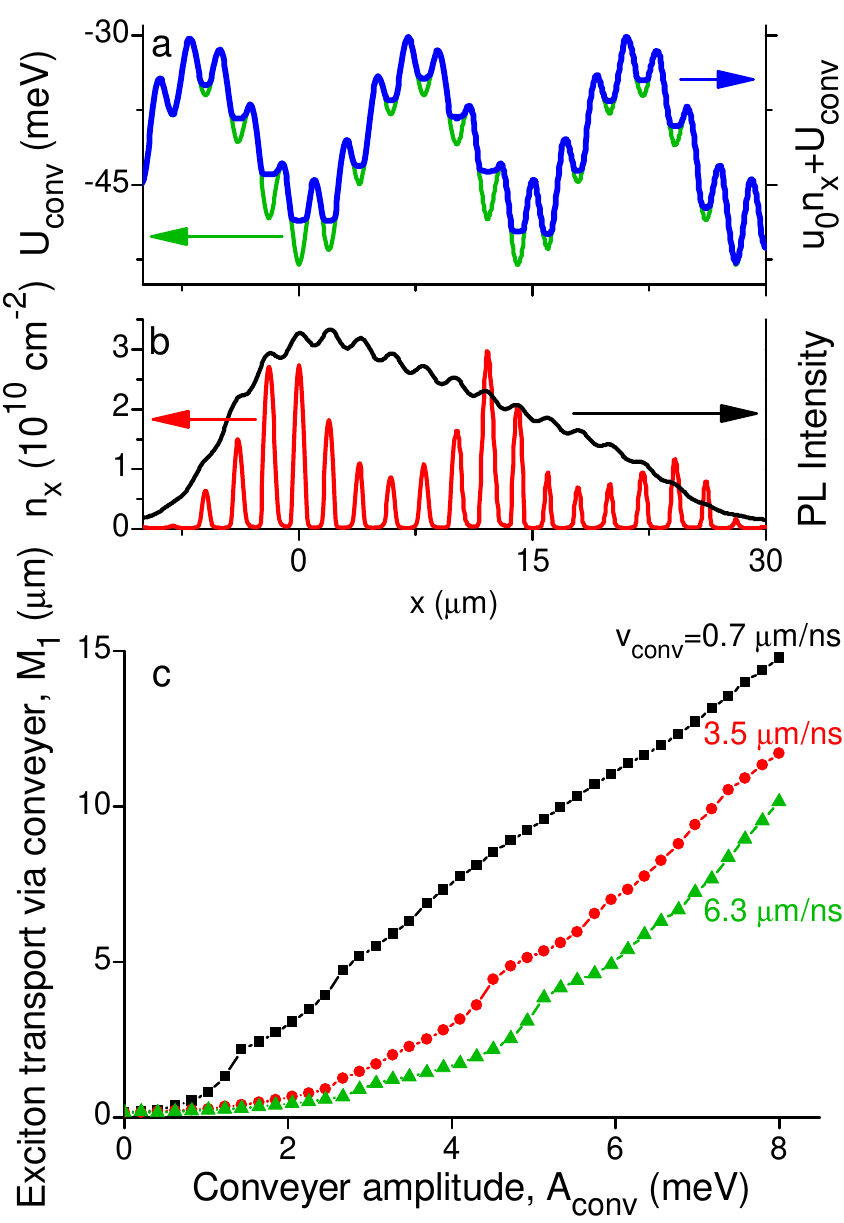}
\caption{Theoretical simulations. (a,b) A snapshot of the conveyer potential (green), exciton density distribution (red), and corresponding screened conveyer potential (blue). PL intensity (black). $A_{\rm conv}=8$ meV, $v_{\rm conv} = 0.7 \mu$m/ns. (c) $M_1$ vs $A_{\rm conv}$. $D_{\rm x}^{(0)}=30{\rm cm^2s^{-1}}$, $\Lambda = 10^9{\rm cm^{-2} ns^{-1}}$, $\tau_{\rm opt}=50{\rm ns}$.}
\end{center}
\end{figure}

The results of the simulations are presented in Fig. 4. The green curve in Fig. 4a presents a snapshot of the conveyer potential for $n_{\rm x}=0$. The sinusoidal envelop of the conveyer potential with $\lambda_{\rm conv} = 14 \mu$m is modulated by $2 \mu$m-period ripples, which originate from the finite spacing between the conveyer electrodes. The amplitude of these ripples can be controlled by the spacing between the conveyer electrodes $d_{\rm s}$ (the ripples essentially vanish for $d_{\rm s} < 0.5 \mu$m for the structure). The repulsively interacting indirect excitons screen the external potential \cite{Ivanov2002}. The snapshot of the exciton density distribution is shown by the red curve and the corresponding screened conveyer potential by the blue curve in Fig. 4a,b.

The time-integrated exciton PL with the spatial resolution taken into account is presented by the black curve in Fig. 4b. The average exciton transport distance via conveyer is evaluated by $M_1$, excluding the shaded area shown in Fig. 1g,h as for the experimental data. The obtained theoretical simulations of exciton transport via conveyer (Fig. 4c) are in qualitative agreement with the experimental data (Fig. 2) exhibiting the dynamical exciton delocalization in shallow conveyers and the dLDT with increasing conveyer amplitude. The simulated and measured conveyer amplitude at the dLDT, $A_{\rm turn-on}$, are in qualitative agreement (Fig. 2b).

In order to simplify the calculations for the analysis of the exciton density dependence, we approximate the conveyer potential by a cosine function $U_{\rm conv}^* = \Delta + A_{\rm conv} {\rm cos}\left( 2\pi(x / \lambda_{\rm conv} - f_{\rm conv}t) \right)$ and and treat the ripples in the same way as the disorder potential. The position of the ripples is fixed, similar to the position of the CQW disorder potential. We approximate the effect of the CQW disorder and conveyer ripples on exciton transport within the thermionic model \cite{Ivanov2002} via the modification of the exciton diffusion coefficient $D_{\rm x} = D_{\rm x}^{(0)}{\rm exp}\left(-(U^{(0)} + U^{(0)}_{\rm ripple})/(k_{\rm B}T + u_{\rm 0}n_{\rm x})\right)$. Here $U^{(0)}_{\rm ripple}$ is the ripple amplitude obtained by simulations. $U^{(0)}_{\rm ripple}$ is nearly proportional to $U_{\rm conv}^*$ and, therefore, it is approximated by $U^{(0)}_{\rm ripple} = C U_{\rm conv}^*$ ($C$ is a fitting constant). The simulated density dependence of exciton transport via conveyer is in qualitative agreement with the experimental data (Fig. 3). The results are discussed below.

{\it Conveyer amplitude dependence (Figs. 2a, 4c).} When the conveyer amplitude is smaller than the exciton interaction energy or disorder amplitude, excitons are not localized in the minima of the moving conveyer potential, in analogy to the case of static lattices \cite{Remeika2009}, and therefore are not moved by the conveyer. When the conveyer amplitude becomes larger than both the exciton interaction energy and disorder amplitude, excitons can localize in the minima of the moving conveyer potential. This results in efficient transport of excitons via conveyer. The effect of the ripples in conveyer potentials on exciton transport is similar to that of disorder. More efficient exciton transport can be achieved by reducing the ripple amplitude. This can be realized by reducing $d_{\rm s}$. The saturation of $M_1$ at large $A_{\rm conv}$ can be related to a device imperfectness and can be studied in future works.

{\it Conveyer velocity dependence (Figs. 2a,b, 4c).} Excitons can efficiently follow the moving conveyer potential when the maximum exciton drift velocity in the conveyer is higher than the conveyer velocity, $v_{\rm drift} = \mu_{\rm x} (\partial U_{\rm conv} / \partial x)_{\rm max} \gtrsim v_{\rm conv}$. This leads to an estimate $A_{\rm turn-on} \sim v_{\rm conv} \lambda_{\rm conv} / \mu_{\rm x}$, qualitatively showing that a higher conveyer amplitude is required for efficient exciton transport via conveyer at a higher $v_{\rm conv}$.

A monotonic dependence of $A_{\rm turn-on}$ on $v_{\rm conv}$ without abrupt changes at the sound velocity is consistent with the thermal velocity of excitons, $\sqrt{2k_{\rm B}T/M_{\rm x}} \sim 15 \mu$m/ns at $T = 1.7$ K, being much higher than the sound velocity.

{\it Density dependence (Fig. 3).} At low densities, the excitons are localized in local minima of the disorder potential (given by the intrinsic disorder and ripples in the conveyer potential) and hardly follow the moving conveyer. At the intermediate densities, excitons effectively screen the disorder and can be efficiently moved by the conveyer. Exciton transport via conveyer becomes less efficient at the high densities when excitons screen the conveyer potential. The requirement for efficient exciton transport via conveyers $\mu_{\rm x} (\partial U_{\rm conv} / \partial x)_{\rm max} \gtrsim v_{\rm conv}$ is relevant, where screening of disorder results in the enhancement of $\mu_{\rm x}$ while screening of the conveyer potential results in the reduction of the conveyer amplitude.

In summary, we report on the realization of electrostatic conveyers for excitons and experimental and theoretical studies of exciton transport via conveyers.

In memory of Alexei Ivanov.

We thank Misha Fogler, Nikolai Gippius, and Egor Muljarov for discussions. This work was supported by the DOE Office of Basic Energy Sciences under award DE-FG02-07ER46449. The development of the conveyer RF system and multilayer lithography was also supported by ARO under award W911NF-08-1-0341 and NSF under award 0907349. Cardiff group was supported by EPSRC, CUROP and WIMCS. This work was performed using the computational facilities of the ARCCA Division, Cardiff University.

\end{document}